\documentclass[english,aps,prl,twocolumn,amsmath,amssymb,superscriptaddress]{revtex4-1}
\usepackage[latin9]{inputenc}
\usepackage{amstext}
\usepackage{amssymb}
\usepackage{graphicx}
\usepackage{color}
\usepackage{babel}

\makeatletter

\providecommand{\tabularnewline}{\\}

\makeatother

\begin{document}


\title{Quantum gas mixtures in different correlation regimes}

\author{Miguel Angel Garcia-March}
\email{magmarch@phys.ucc.ie}
\affiliation{Physics Department, University College Cork, Cork, Ireland}
\author{Thomas Busch}
\affiliation{Physics Department, University College Cork, Cork, Ireland}
\affiliation{Quantum Systems Unit, Okinawa Institute of Science and
  Technology, Okinawa, Japan}

\begin{abstract}
  We present a many-body description for two-component ultracold
  bosonic gases when one of the species is in the weakly interacting
  regime and the other is either weakly or strongly interacting. In
  the one-dimensional limit the latter case system is a hybrid in
  which a Tonks-Girardeau gas is immersed in a Bose-Einstein
  condensate, which is an example of a new class of quantum system
  involving a tunable, superfluid environment. We describe the
  process of phase separation microscopically and semiclassically in
  both situations and show that the quantum correlations are maintained in
  the separated phase.
\end{abstract}

\maketitle

\emph{Introduction} -- Impurities immersed in ultracold atomic gases
have recently emerged as versatile environments for studying quantum
correlations in highly controllable systems~\cite{Cote:02}.
Demonstrations of single fermions or single ions embedded in a Bose
Einstein condensates (BECs) have shown that such systems are
experimentally viable~\cite{Zipkes:10}, and are paving the way to
study a plethora of new quantum phenomena, arising from the
interactions between the impurity and the ultracold
environment. Better understanding and control of these interactions
are already leading to new ideas in quantum information
theory~\cite{Doerk:10}.

Precursors to these highly controllable hybrid systems have been
mesoscopic mixtures of ultracold bosonic gases, which consist of
either two different atomic species or two different hyperfine states
of the same species \cite{Myatt:97}. They are commonly available in
labs worldwide and have allowed to investigate mesoscopic quantum
dynamics in complex systems and to study new exotic states of
matter. Such systems have a successful microscopic description based
on a two mode model \cite{Law:97} and can be approximated
semiclassically by using a system of coupled Gross-Pitaevskii
equations (GPE)~\cite{Ho:96}. The latter model allows to describe the
stability of the multi-component system~\cite{Busch:97} and the
process of phase separation~\cite{Ho:96, Ao:98}.

In this work we consider a small number of atoms confined in an
effectively one-dimensional parabolic trap, which are either in the
weakly or in the strongly correlated regime. Because of the reduced
dimensionality, the strongly interacting case corresponds to the
Tonks-Girardeau (TG) limit~\cite{Girardeau:60, Girardeau:01}, which
has recently been demonstrated experimentally~\cite{Paredes:04}. The
same trap contains a second species of atoms, which we consider to be
in the weakly correlated regime (the BEC limit), and which acts as a
tunable environment for the first component. We show that the quantum
correlations can be tuned through the coupling with between the two
species, and describe microscopically the phase separation process
that drives the immersed species to the edges of the BEC.

A strongly interacting quantum gas in one-dimension can be
successfully described in different ways. The first is to use a model
of hard sphere bosons and employ a mapping theorem to a
non-interacting Fermi gas~\cite{Girardeau:60}. This permits one to
derive an analytical expression for the wave function in position
space and only relies on the knowledge of the solutions of the single
particle problem~\cite{Girardeau:60, Girardeau:01, Goold:10.2}. An
equal, but numerically expensive, description is to use a many-body
Hamiltonian and expand the field operators into a sufficient number of
momenta~\cite{Deuretzbacher:07}.

Here we choose the second approach and expand the second quantized
field operators for both species in a basis of harmonic oscillator
functions, including as many momenta as needed. This allows us to
microscopically describe the stationary solutions and the phenomenon
of phase separation when both species are in the weakly interacting
regime (the BEC-BEC limit) and when one is in the strongly interacting
regime (the BEC-TG limit). While a larger number of momenta is
necessary to describe the TG gas, only a few momenta are necessary for
the BEC species, which makes the numerical approach possible.  We
obtain the many body version of the phase separation criteria and show
that this process occurs as a consequence of the excitation of atoms
to higher harmonics. The phase separated situation can be treated as a
double well potential for the phase separated species and we show that
the atoms in the strongly interacting species keep their correlations
despite being separated by the BEC component.

Since the above approach is limited to small particle numbers, we
generalise the obtained results to the mesoscopic limit by using a
semiclassical approach similar to the well known coupled GPE
systems~\cite{Ho:96}. In the BEC-TG limit, however, we show that a GPE
coupled to a mean field equation with a quintic~\cite{Kolomeisky:00}
rather than a cubic non-linearity appropriately describes the
semiclassical limit, while the non-linearity of the cross term depends
on the value of the coupling between the species.

\emph{Model} -- We consider a mixture of two ultracold bosonic
species, confined in a quasi-one dimensional (1D) parabolic trap. One
component (the 'environment', $E$) is always in the weakly correlated
regime, whereas the other component (the 'system', $S$) can be either
weakly or strongly interacting. The intra-species coupling constants
are then given by $g_{E,S}=-2\hbar^{2}/m_{E,S}a_{1D}^{E,S}$, where
$m_{E,S}$ are the respective masses of the two components and
$a_{1D}^{E,S}=(-(a_{\perp}^{E,S})^{2}/2a_{E,S})[1-C(a_{E,S}/a_{\perp}^{E,S})]$.
Here $a_{E,S}$ are the $s$-wave scattering lengths and
$a_{\perp}^{E,S}=\sqrt{2\hbar/m_{E,S}\omega_\perp}$, where
$\omega_\perp$ is the trap frequency in the radial direction. The
constant $C$ is evaluated in~\cite{Olshanii:98}.  For simplicity we
assume that the two components are different hyperfine states of the
same atomic species and therefore $m_{E}=m_{S}=M$. The coupling
constant between the two components, $ g_{ES}$, is given in a similar
fashion in terms of the 1D inter-species scattering length,
$a_{1D}^{ES}$.  We expand the field operators in the second-quantized
description of this system~\cite{Law:97} in terms of the
eigenfunctions $\phi_{n}(x)$ of the harmonic potential
$V(x)=\frac{1}{2}M\omega^{2}x^{2}$ and use $n_{E,S}$ modes in the
expansion for the environment and system component,
$\hat{\psi}_E(x,t)=\sum_{n=1}^{n_E}a_n(t)\phi_n(x)$ and
$\hat{\psi}_S(x,t)=\sum_{n=1}^{n_S}b_n(t)\phi_n(x)$. The creation and
annihilation operators $a_{k}^{\dagger}$ and $a_{k}$ satisfy the
standard (equal time) bosonic commutation relations
$[a_{k},a_{l}^{\dagger}]=\delta_{kl},$
$[a_{k},a_{l}]=[a_{k}^{\dagger},a_{l}^{\dagger}]=0$ (and similarly for
$b_{k}^{\dagger}$ and $b_{k}$), which yields the Hamiltonian
$H=H_E+H_S+H_{\mathrm{int}}$, where
\begin{align}
  H_E =&\sum_{k,l}a_k^\dagger a_lH_{kl}+\frac{1}{2}
        \sum_{klmn}a_k^\dagger a_l^\dagger a_ma_nV_{klmn}^E\\
  H_S =&\sum_{k,l}b_k^\dagger b_lH_{kl}+\frac{1}{2}
        \sum_{klmn}b_k^\dagger b_l^\dagger b_mb_nV_{klmn}^S,\\
  H_{\mathrm{int}}=&\frac{1}{2}\sum_{klmn}a_k^\dagger b_l^\dagger b_ma_nV_{klmn}^{ES},
\end{align}
with 
\begin{align}
 H_{kl}=&\int dx\;\phi_k^*(x)H_\mathrm{sp}\phi_l(x),\\
 V_{klmn}^{E,S}=&
         \frac{g_{E,S}}{2}\int dx\;\phi_k^*(x)\phi_l^*(x)\phi_m(x)\phi_n(x),\\
 V_{klmn}^{ES}=&g_{ES}\int dx\;\phi_k^*(x)\phi_l^*(x)\phi_m(x)\phi_n(x).
\end{align}
Here $H_\mathrm{sp}$ is the single particle Hamiltonian for the
harmonic oscillator. We then expand the ground state
$\Psi_{0}=\sum_{i=1}^{\Omega}c_{i}\Phi_{i}$ as a sum over all Fock
vectors given by
\begin{equation}
 \Phi_i=D\left(a_1^\dagger\right)^{N_1^E}\!\!\!\!\dots
         \left(a_{n_E}^\dagger\right)^{N_{n_E}^E}
         \left(b_1^\dagger\right)^{N_1^S}\!\!\!\!\dots
         \left(b_{n_S}^\dagger\right)^{N_{n_S}^S}\Phi_0,
\end{equation}
with $D=(N_{1}^{E}!\dots N_{n_{E}}^{E}!N_{1}^{S}!\dots
N_{n_{S}}^{S}!)^{-\frac{1}{2}}$ and $\Phi_{0}$ being the vacuum. Here
$N_{1}^{E},\dots,N_{n_{E}}^{E}$ ($N_{1}^{S},\dots,N_{n_{S}}^{S}$) are
the occupation numbers of the $n_{E}$ ($n_{S}$) modes for the
environment (system). The dimension of the Hilbert space is
$\Omega=\Omega_{E}\Omega_{S}$ with
$\Omega_{E,S}=(N_{E,S}+n_{E,S}-1)!/N_{E,S}!(n_{E,S}-1)!$ where
$N^{E,S}$ is the total number of atoms in each species. The fast
growth of this space for larger particle numbers or modes is the
biggest challenge to the numerical approach.

\begin{figure}
\begin{tabular}{cc}
\includegraphics[scale=0.35]{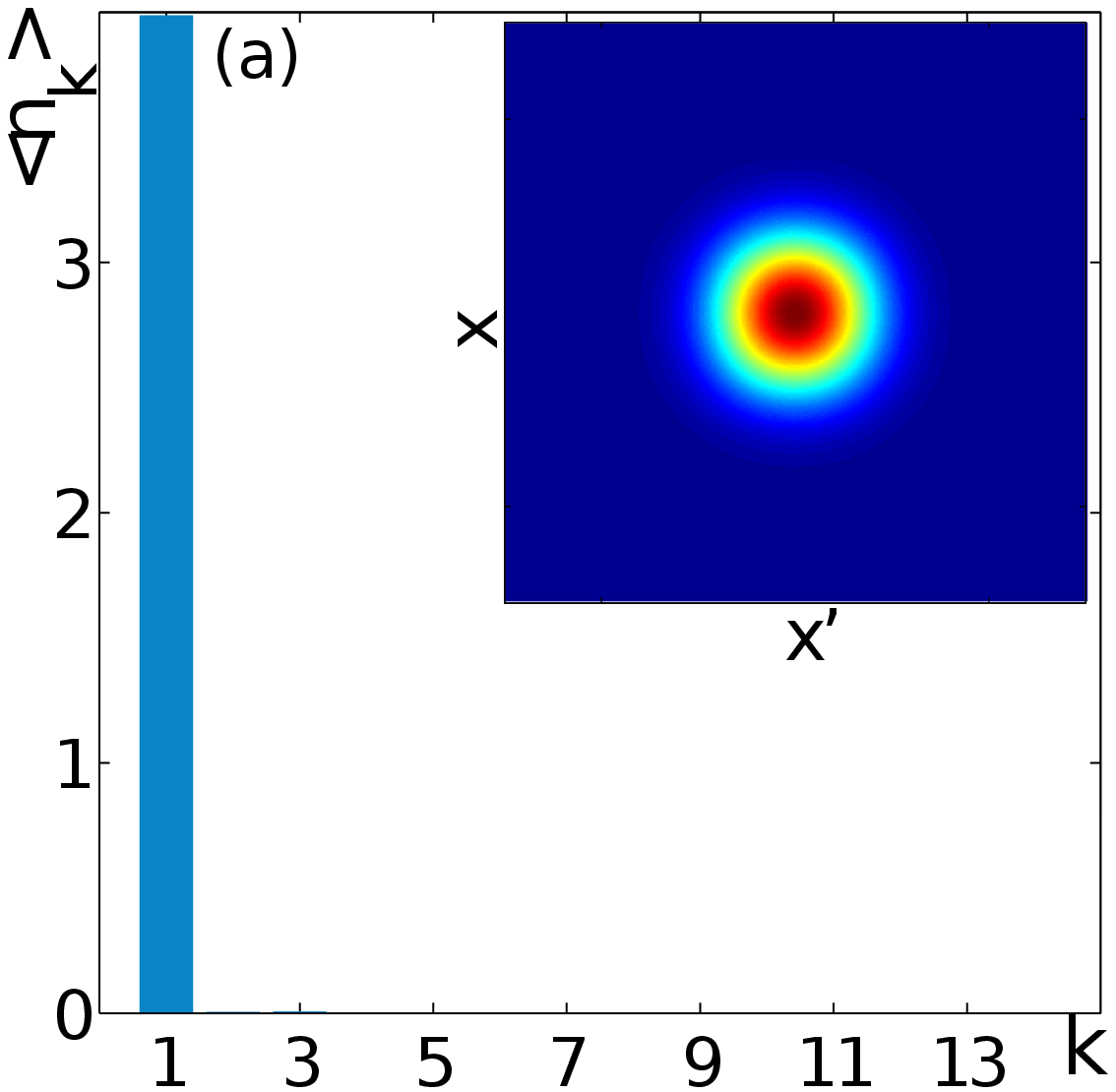} & \includegraphics[scale=0.27]{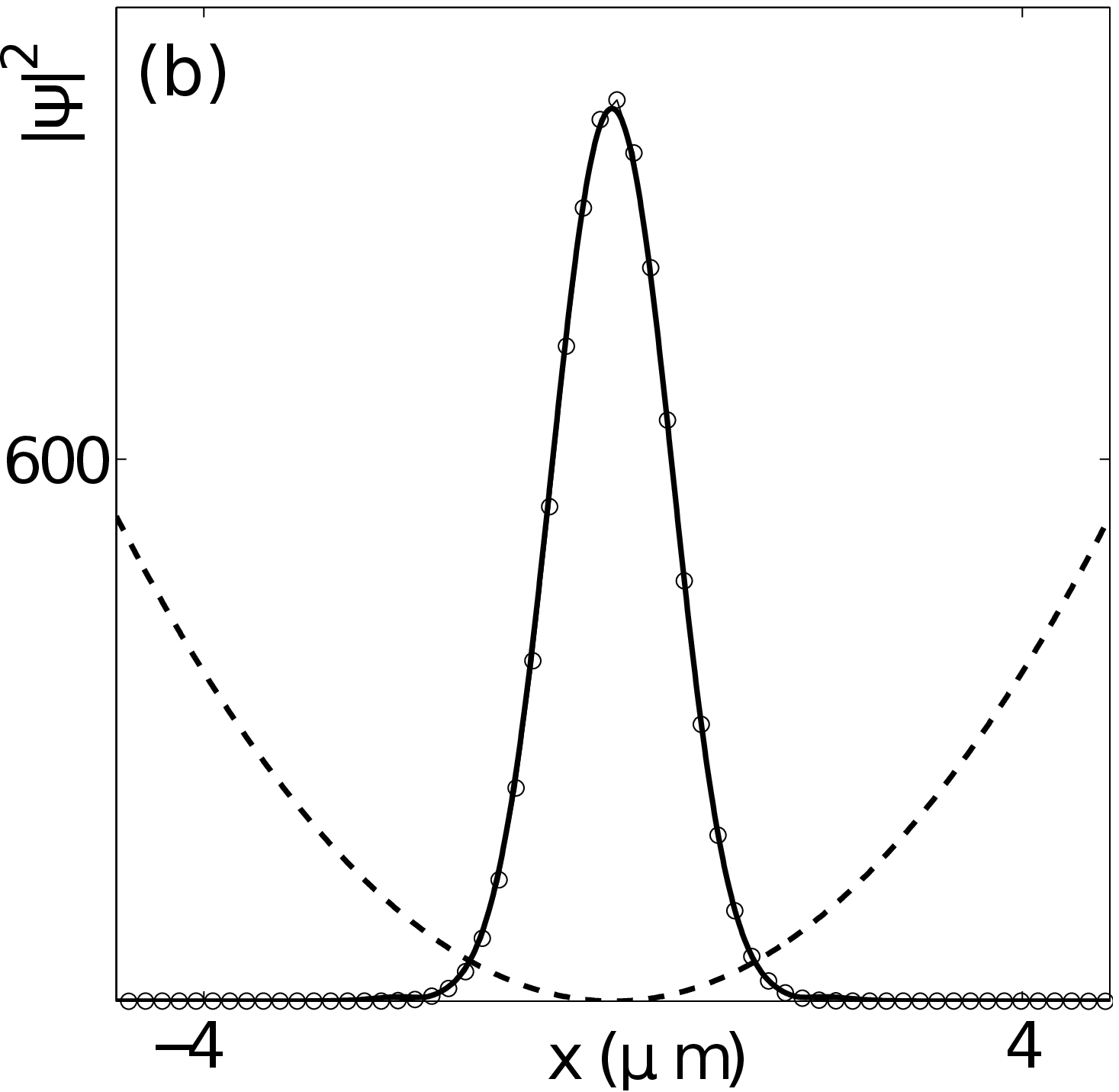}\tabularnewline
\includegraphics[scale=0.35]{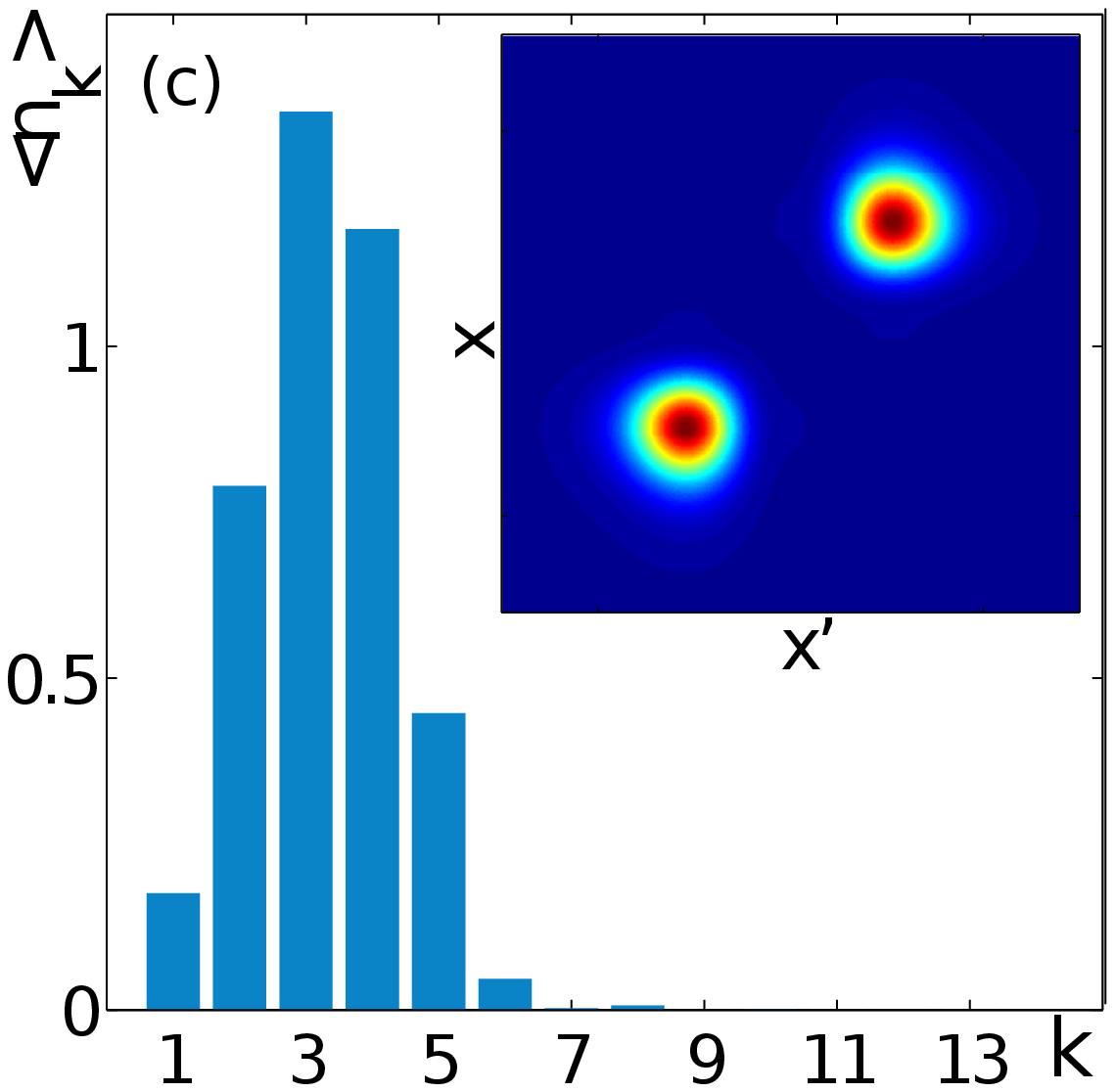} & \includegraphics[scale=0.27]{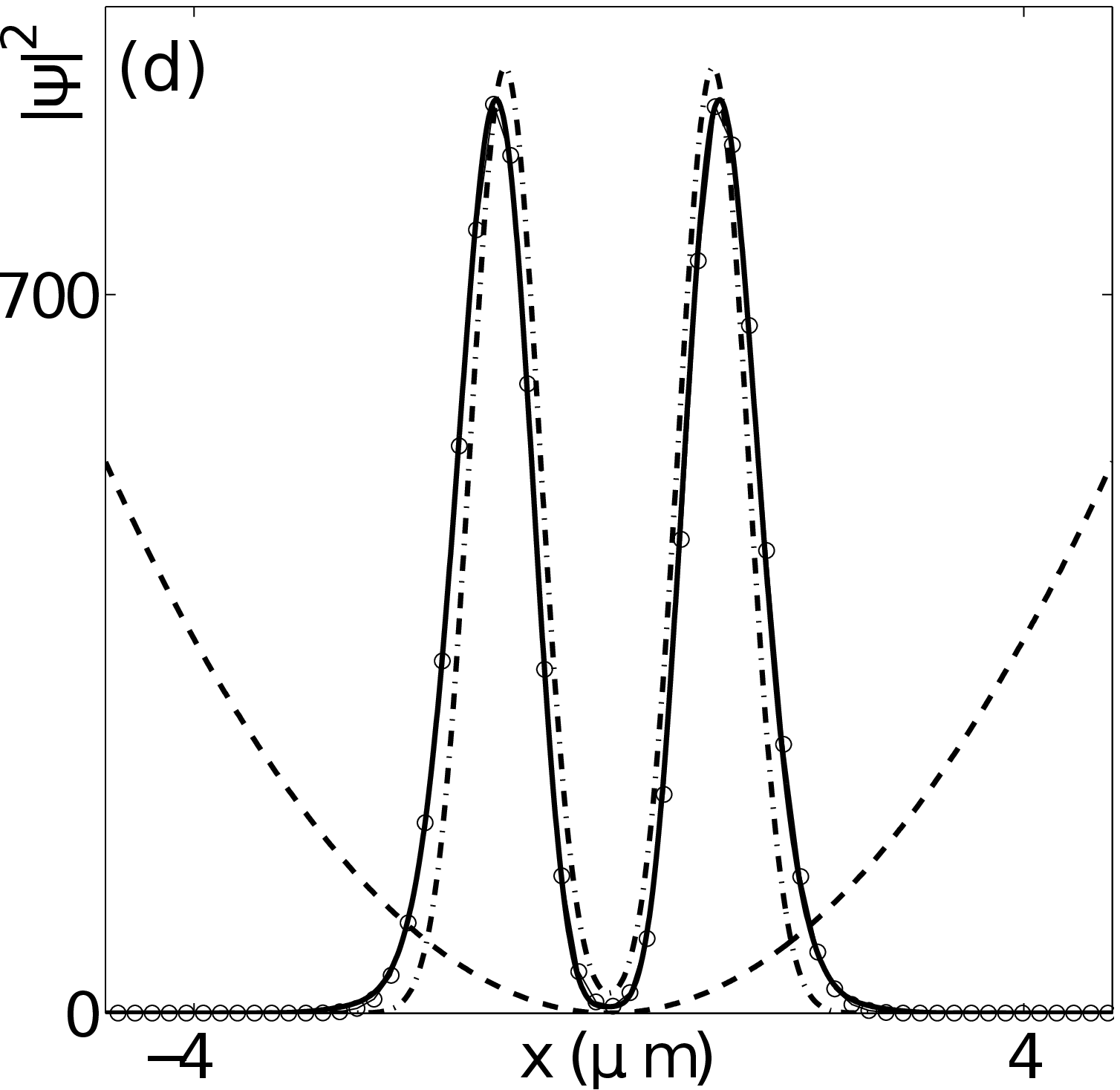}
\end{tabular}
\vspace{-0.4cm}
\caption{\label{fig:Fig1}BEC-BEC regime without (upper row) and with
  (lower row) coupling between species.  The left column shows the
  average occupations $\left\langle n_{k}\right\rangle $ of the
  momenta $k$. Insets are the corresponding SPDMs, spanning a region
  of $6\mu m\times6\mu m $.  The right column shows the density
  profiles, with the dashed line being the potential, the solid line
  being the density obtained from the SPDM and the solid line with
  circled markers being the one obtained from a semiclassical GPE-GPE
  simulations. The dash-dotted lines in (d) are the approximate
  Gaussians located at the minima of the double well potential
  $V_{NL}(x)=V(x)+g_{AB}N_{B}|\phi_{0}(x)|^{2}|$.  }\vspace{-0.5cm}
\end{figure}

\emph{BEC-BEC regime} -- If the interactions $g_{E,S}$ are small and
$g_{ES}$ vanishes it is sufficient to use only a few modes to describe
both species. Indeed, if the intra-species coupling constants vanish
as well, the single particle density matrix (SPDM), defined as
$\rho_{E}(x,x')=\sum_{k,k'}\phi_{k}(x)\phi_{k'}(x')\left\langle
\right.\!\! a_{k}^{\dagger}a_{k'}\!\!\left.\right\rangle $ and
similarly for $\rho_{S}(x,x')$, is simply a Gaussian of width
$\alpha=\sqrt{\hbar/M\omega}$.  For weak interactions the atoms still
mainly occupy the lowest energy eigenfunction ( $N_{1}^{E,S}\approx
N^{E,S}$) and a Gaussian approximation to the eigenstates will be
good. Assuming spatial overlap between both components, the energy of
the ground state $\Psi_{0}$ can then be found as
\begin{align}
  \label{eq:Ehom_BECBEC}
  E^\mathrm{hom}=&(N_E+ N_S)\frac{\hbar\omega}{2}
                +\frac{g_E}{2\sqrt{2\pi}\alpha}N_E(N_E-1)\nonumber\\
               &+\frac{g_S}{2\sqrt{2\pi}\alpha}N_S(N_S-1)
                +\frac{g_{ES}}{\sqrt{2\pi}\alpha}N_{E}N_{S},
\end{align}
where we have used that
$V_{0000}^{E,S}=g_{E,S}/2\sqrt{2\pi}\alpha$. If the two species phase
separate, one of them, say $E$, will stay in the center of the trap,
approximately with a Gaussian shape of width $\alpha_{E}$. A single
Gaussian mode then suffices in the expansion of the field operator of
the environment. The system component will approximately assume a
profile formed by two Gaussians of width $\frac{\alpha_S}{2}$ (see
Fig.~\ref{fig:Fig1}(d)), which we call $\varphi_{0}^{DW}$. In the
harmonic oscillator basis the expansion of this state requires a
larger number of modes, however, if we use the approximation of
$\varphi_{0}^{DW}$ as two displaced Gaussians in a potential of
frequency $\omega'$, we find the energy of this state as
\begin{align}
\label{eq:Einh_BECBEC}
  E^\mathrm{inh}=&N_E\frac{\hbar\omega}{2}
                  +\frac{g_E}{2\sqrt{2\pi}\alpha_E}N_E(N_E-1)\nonumber\\
               &+N_S\frac{\hbar\omega'}{2}
                  +\frac{g_S}{2\sqrt{2\pi}\alpha_S}N_S(N_S-1).
\end{align}
To determine the point at which phase separation happens, we
assume that the atoms occupy the same total volume before and after
phase separation, $\alpha_{E}+\alpha_{S}=\alpha$ \cite{Ao:98}, and
minimize the energy with respect to $\alpha_E$ and $\alpha_S$. This
gives $\alpha_E=\frac{\alpha}{1+k}$ and
$\alpha_S=\frac{\alpha}{1+k^{-1}}$, with
$k=\sqrt{\frac{g_S}{g_E}\frac{N_S(N_S-1)}{N_E(N_E-1)}}$.  Subtracting
both energies  we obtain the
phase separation criterion 
\begin{equation}
 \label{eq:PhaseSeparation}
 g_{ES}N_EN_S>\sqrt{g_Eg_S}\sqrt{N_E(N_E-1)N_S(N_S-1)}, 
\end{equation}
where we have assumed that the change in kinetic energy $\hbar
N_S(\omega'-\omega)$ is negligible compared to the overall change in
interaction energies (we justify this assumption below). This
criterion resembles the semiclassical one for large particle numbers,
$N_{E,S}(N_{E,S}-1)\approx N_{E,S}^{2}$~\cite{Ho:96, Ao:98}. Note that
it predicts phase separation for any nonzero value of $g_{ES}$ if one
of the components consists only of a single particle.  If we expand
$\varphi_0^{DW}$ in terms of the eigenfunctions of the harmonic
potential $\varphi_{0}^{DW}=\sum\left\langle
  \varphi_{0}^{DW}|\phi_{k}\right\rangle \phi_{k}$, we can see that
the process of phase separation coincides with the occupation of more
and more orbitals in the harmonics basis. Consequently, the phase
separated species will be the one with the higher coupling constant
and the smaller number of atoms and for simplicity we therefore assume
in the following $g_S> g_E$ and $N_E\gg N_S$. This assumption also
makes numerical calculations possible, since the shape of the
environment component does not change substantially after phase
separation ($\alpha_E\approx \alpha$) and it can correspondingly be
described only considering a single mode.

Let us consider an example using atoms of the mass of $^{87}$Rb in a
trap of frequencies $\omega=2\pi\cdot 400$Hz and
$\omega_\perp=100\omega$. We (exemplary) choose the scattering lengths
to be $a_E= a_0/10$ and $a_S=10 a_0$, where $a_0$ is the Bohr
radius. The environment consists of $N_{E}=300$ particles and the
immersed component of $N_{S}=4$, giving $g_E=10^{-9}\hbar\omega$ and
$g_S=1.01\times 10^{-7}\hbar\omega$. In all calculations we use
$n_S=14 $ and $n_E=1$ (as justified above). In the upper row in
Fig.~\ref{fig:Fig1} we show the situation where the different species
do not interact ($g_{ES}=0$), while in the bottom row
$g_{ES}=\frac{g_S}{2}$, which is deep inside the phase separated
regime.  As expected, for $g_{ES}=0$ the SPDM of both species is
Gaussian, as shown in the inset of Fig.~\ref{fig:Fig1}(a) for the
system component. The energy of the sample in this case is given by
eq.~\eqref{eq:Ehom_BECBEC} as $E^{\mathrm{hom}}=
187.0\,\hbar\omega$. Correspondingly, the average occupation for the
system component $\langle n_k\rangle=\langle b_k^\dagger b_k\rangle$,
represented in Fig.~\ref{fig:Fig1}(a), shows that mainly one momentum
component is occupied, because the coupling constant is very
small. For bigger $g_S$ higher lying momenta will start gaining
occupation, however the system will still form a BEC as the occupation
of the lowest momentum component is of the order of $N_S$. As long as
the sample is in this limit no phase separation
occurs. Fig.~\ref{fig:Fig1}(b) shows the density of the system
component obtained from the SPDM (solid line) coinciding with the
semiclassical calculation using two coupled GPE (solid line with
circled markers).

If $g_{ES}=\frac{g_S}{2}$ (lower row in Fig.~\ref{fig:Fig1}) the
energy for a phase separated state is much smaller than that for a
mixed phase,
$E^{\mathrm{inh}}=191.0\,\hbar\omega<E^{\mathrm{hom}}=230.8\,\hbar\omega$
and also corresponds approximately to the one calculated numerically
of $E=195.0\,\hbar\omega$.  Accordingly, the SPDM in this situation
(see inset of Fig.~\ref{fig:Fig1}(c)) shows that phase separation has
occurred and an occupation of higher momentum modes is found (see
Fig.~\ref{fig:Fig1}(c)).  The densities calculated from the SPDM and
obtained from the numerical solution of the coupled GPE shows again
good agreement (see Fig.~\ref{fig:Fig1}(d)). The Gaussian functions
used to calculate the energy given by eq.~\eqref{eq:Einh_BECBEC} are
located at the minima of the double well potential
$V_{NL}(x)=V(x)+g_{ES}N_S|\phi_0(x)|^2|$, which are given by $d=\pm
\alpha\sqrt{\beta}$, with
$\beta=\ln\left(2g_{ES}N_{S}/\alpha^3M\sqrt{\pi}\omega^2\right)$. The
local trapping frequency can then be approximated as
$\omega'=\sqrt{2}\omega\beta$, leading to a width of the Gaussians of
$\alpha'=\sqrt{\hbar/M\omega'}$. We can now estimate the increase in
the kinetic energy due to phase separation as $ \hbar
N_S\omega(1-\sqrt{2}\beta)$, which has to be compared to the change in
the interaction energies ($g_{ES}N_EN_S$ and
$\sqrt{g_Eg_S}\sqrt{N_E(N_E-1)N_S(N_S-1)}$). Since the latter is
quadratic in the number of particles it is generally much larger
except for systems with very small coupling constants and number of
atoms.

\begin{figure}
\begin{tabular}{cc}
\includegraphics[scale=0.35]{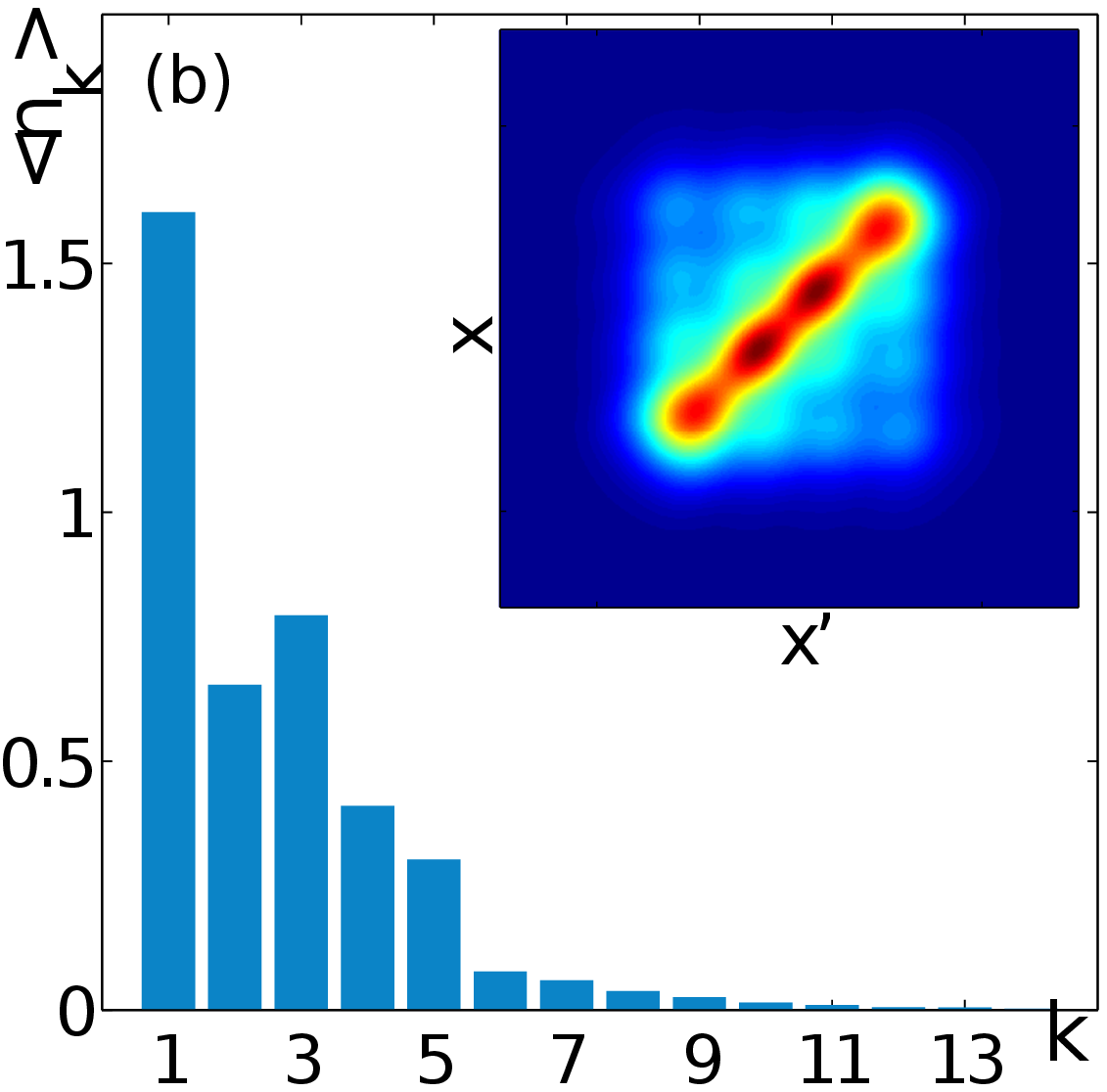} & \includegraphics[scale=0.265]{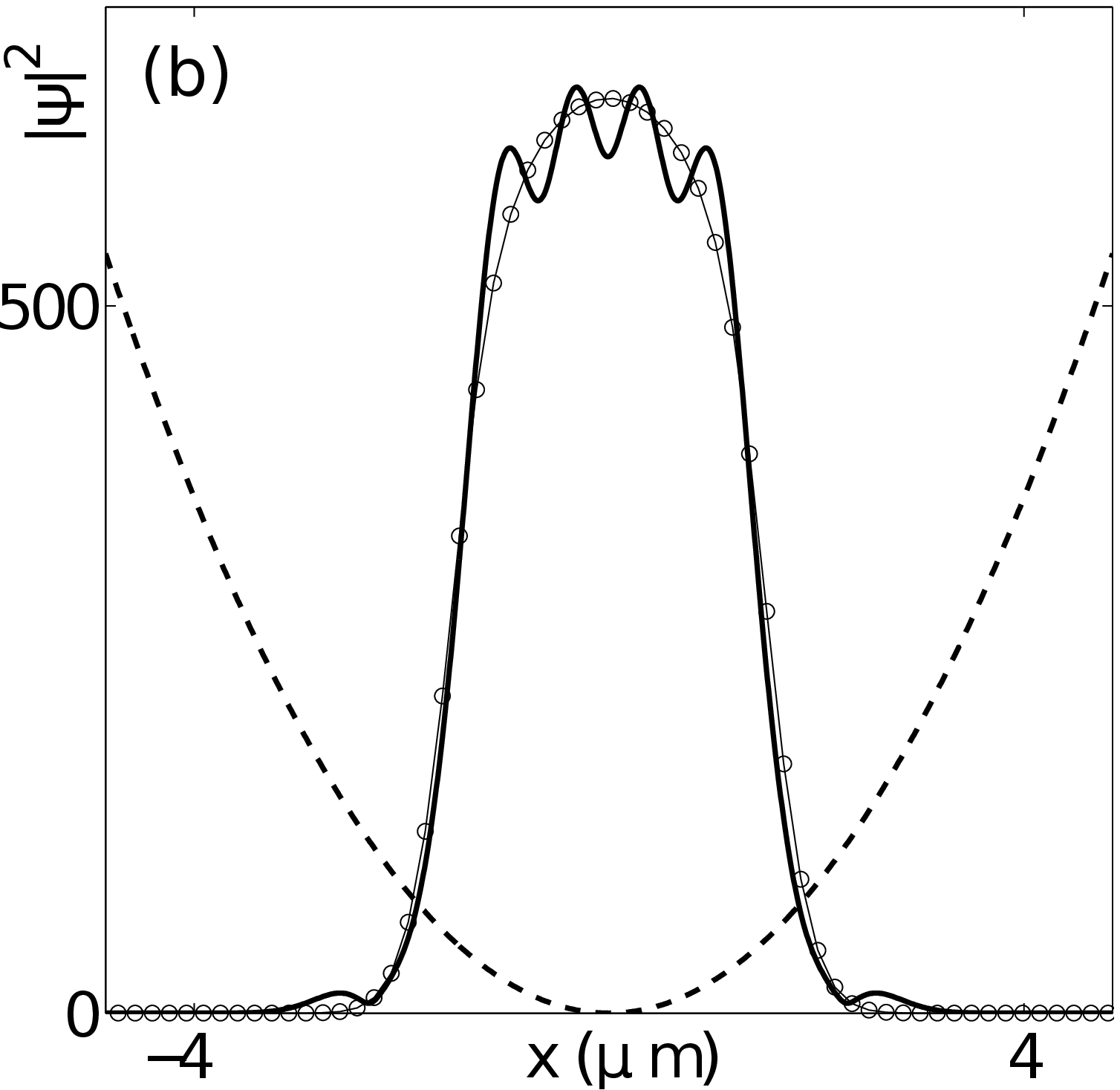}\tabularnewline
\includegraphics[scale=0.35]{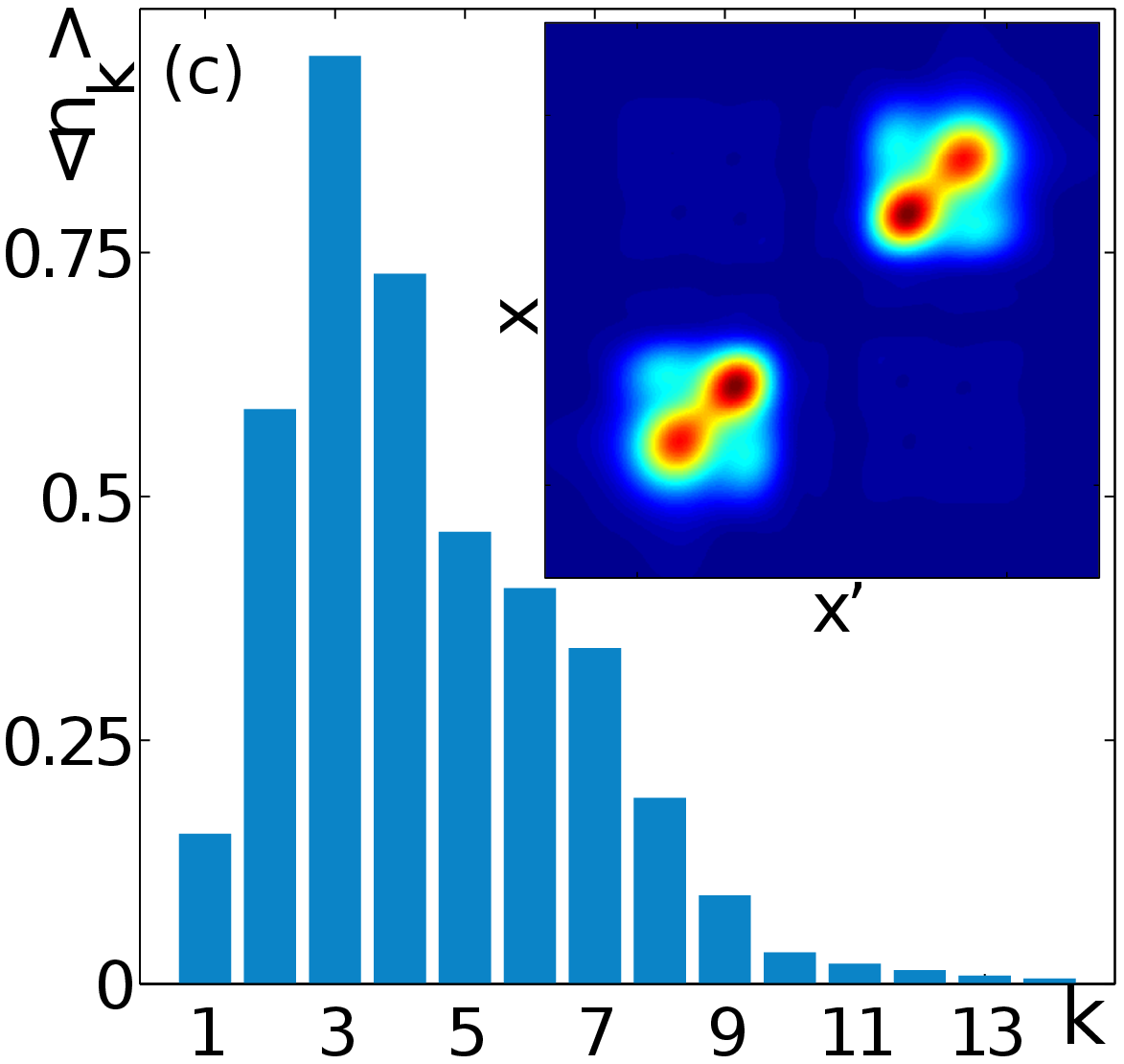} & \includegraphics[scale=0.265]{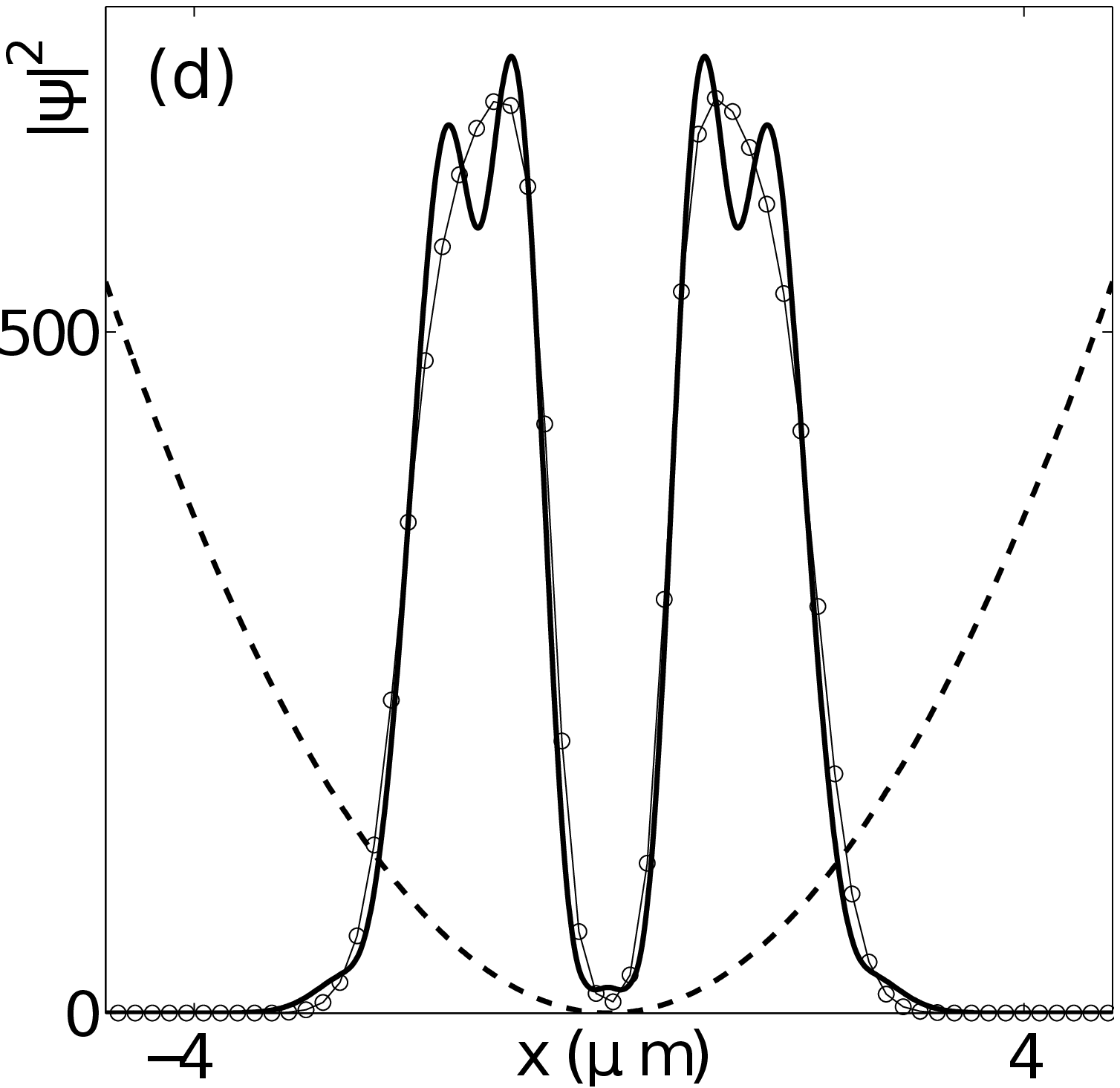}
\end{tabular}\vspace{-0.4cm}
\caption{\label{fig:Fig2} BEC-TG regime without (upper row) and with
  (lower row) coupling between species. Panel lay-out as in
  Fig.~\ref{fig:Fig1}.  The semiclassical profiles are obtained using
  the mean field equations described in the text. }\vspace{-0.5cm}
\end{figure}

\emph{BEC-TG regime} -- We now consider the case where the system
component is in the Tonks-Girardeau regime.  In Fig.~\ref{fig:Fig2} we
show the same quantities as before, but now for $N_E=40$,
$a_{E}=a_{0}$ and $a_{S}=500a_{0}$, which gives a Lieb-Liniger
parameter of $\gamma=2g_{S}ML/\hbar^{2}N_S=4.5$~\cite{Lieb:63}, where
$L$ is the size of the cloud. This corresponds to the system component
being in the Tonks-Girardeau regime and we find the interaction
coefficients to be $g_E=10^{-8}\hbar\omega$ and $g_S=9.6\times
10^{-5}\hbar\omega$. As expected, for no inter-species interaction the
SPDM and the momentum distribution for the system component resemble
that of a TG gas (Fig.~\ref{fig:Fig2} (a)).  The density profiles
calculated from the microscopic and the semiclassical approach (see
below and Fig.~\ref{fig:Fig2} (b)) show again good agreement. Though
the phase separation criterion of eq.~\eqref{eq:PhaseSeparation} is
only approximately valid in this regime, due to the deviations from
the Gaussian shapes for the individual components, it is still useful
away from the precise border transition point.  The lower row of
Fig.~\ref{fig:Fig2} shows the situation for $g_{ES}=0.05g_{S}$, where
phase separation is clearly visible in the SPDM and the momentum
distribution shows a shift of the momenta due to the density
resembling a higher excited state.

To compare the above results with a mean-field model in the
semiclassical limit, we chose to model the strongly correlated system
component using a quintic non-linearity in the field
equation~\cite{Kolomeisky:00}. While one has to be aware that this
approach for a TG gas does not describe the coherence properties
correctly~\cite{Girardeau:00}, it is known to give a good
approximation to the exact density in the single component case. This
equation is then coupled to a GPE for the BEC component, giving
\begin{subequations}
\begin{align}
  i\hbar\dot\psi_S=& -\frac{\hbar^2}{2m}\psi_S^{''}
                 +\left[V(x)+\tilde g_S|\psi_S|^4+g_{ES}|\psi_E|^p\right]\psi_S\\
  i\hbar\dot\psi_E=& -\frac{\hbar^2}{2m}\psi_E^{''}
                 +\left[V(x)+g_E|\psi_E|^2+g_{ES}|\psi_S|^p\right]\psi_E,
\end{align}
\label{KGPE}
\end{subequations}
where $\tilde g_S=\frac{\left(\pi\hbar\right)^{2}}{2M}$.  If the
interactions between both components are small,
i.e.~$\gamma'=2g_{ES}ML/\hbar^{2}N_S<1$, the exponent on the
non-linear coupling term is given by $p=2$ and in the opposite limit
by $p=4$. For the parameters used in Fig.~\ref{fig:Fig2} we find
$\gamma'=0.22$ and the numerical solution of eqs.~\eqref{KGPE} show
good agreement with the microscopically calculated density (see
Fig.~\ref{fig:Fig2}(c)). Calculations for increased values of
$g_{ES}$, which require $p=4$ also show good agreement, but are not
shown here.

\begin{figure}

\begin{tabular}{cc}
\includegraphics[scale=0.35]{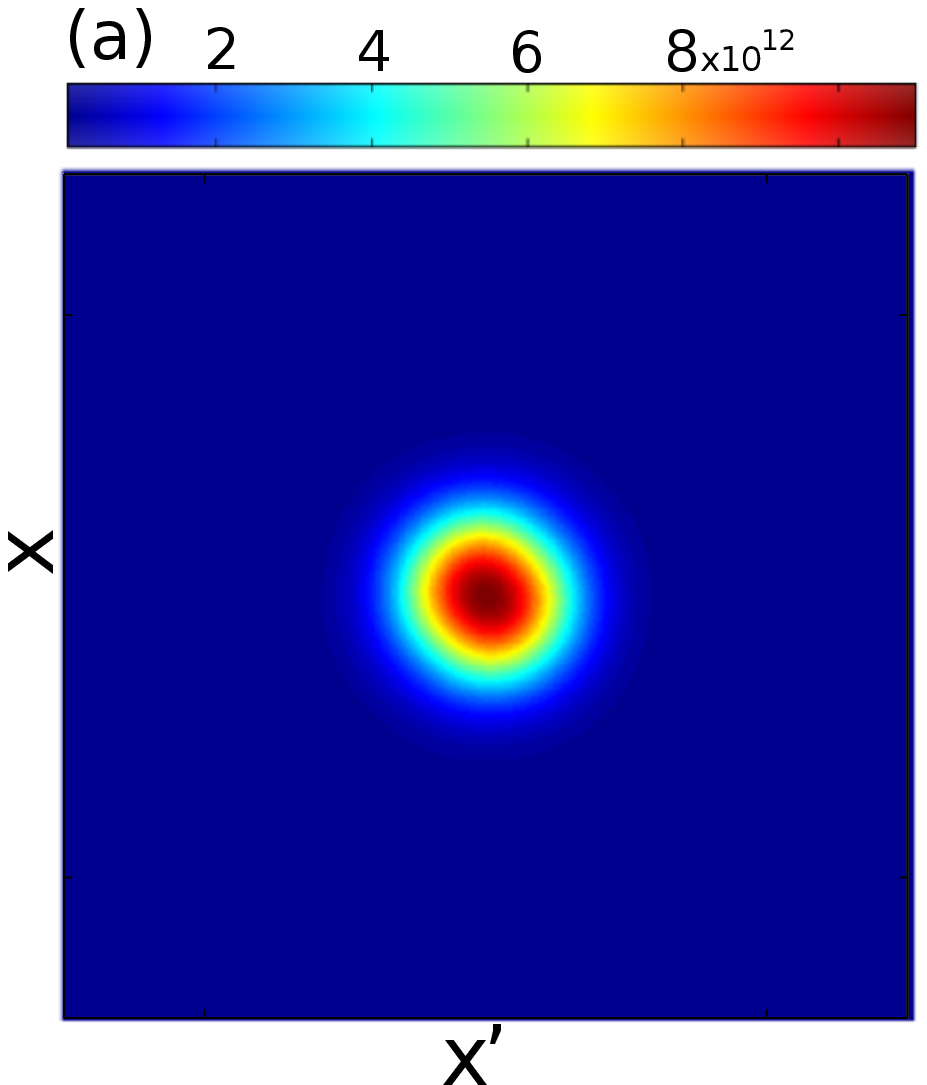} & \includegraphics[scale=0.35]{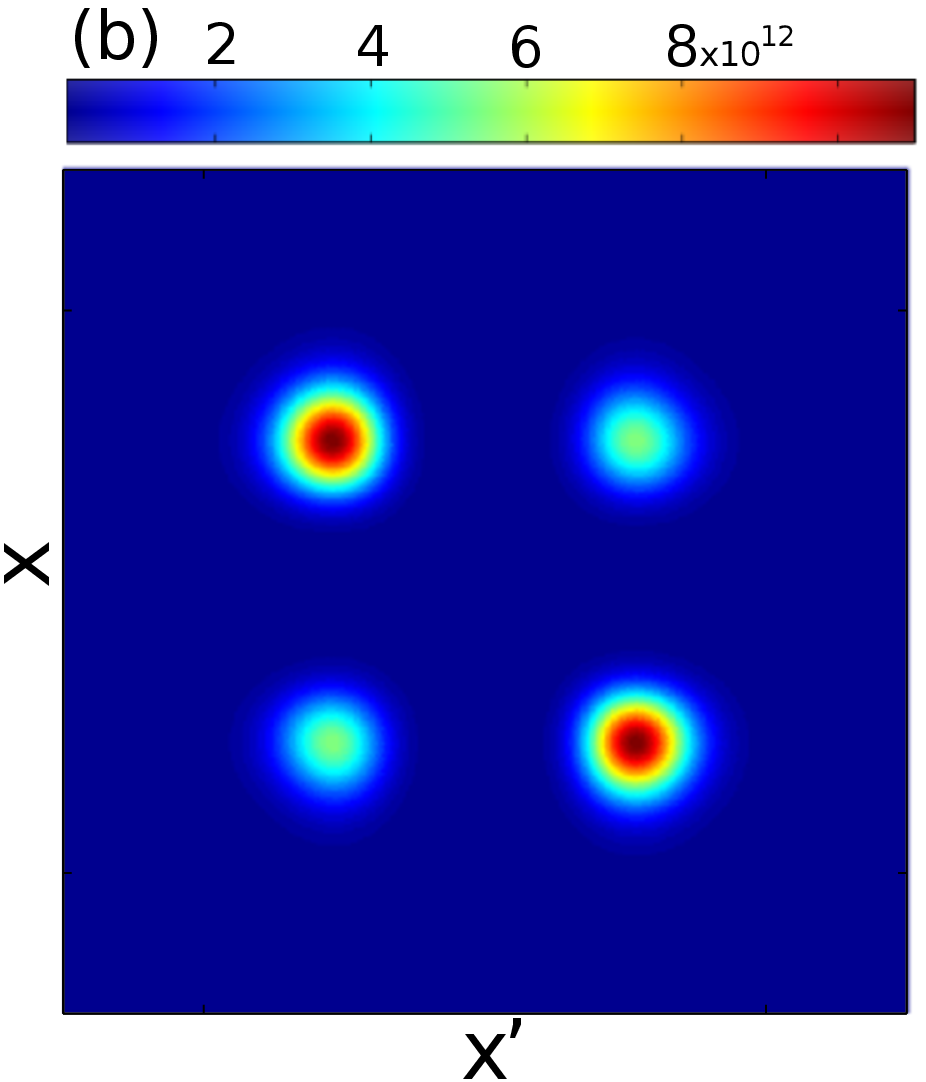}\tabularnewline
\includegraphics[scale=0.35]{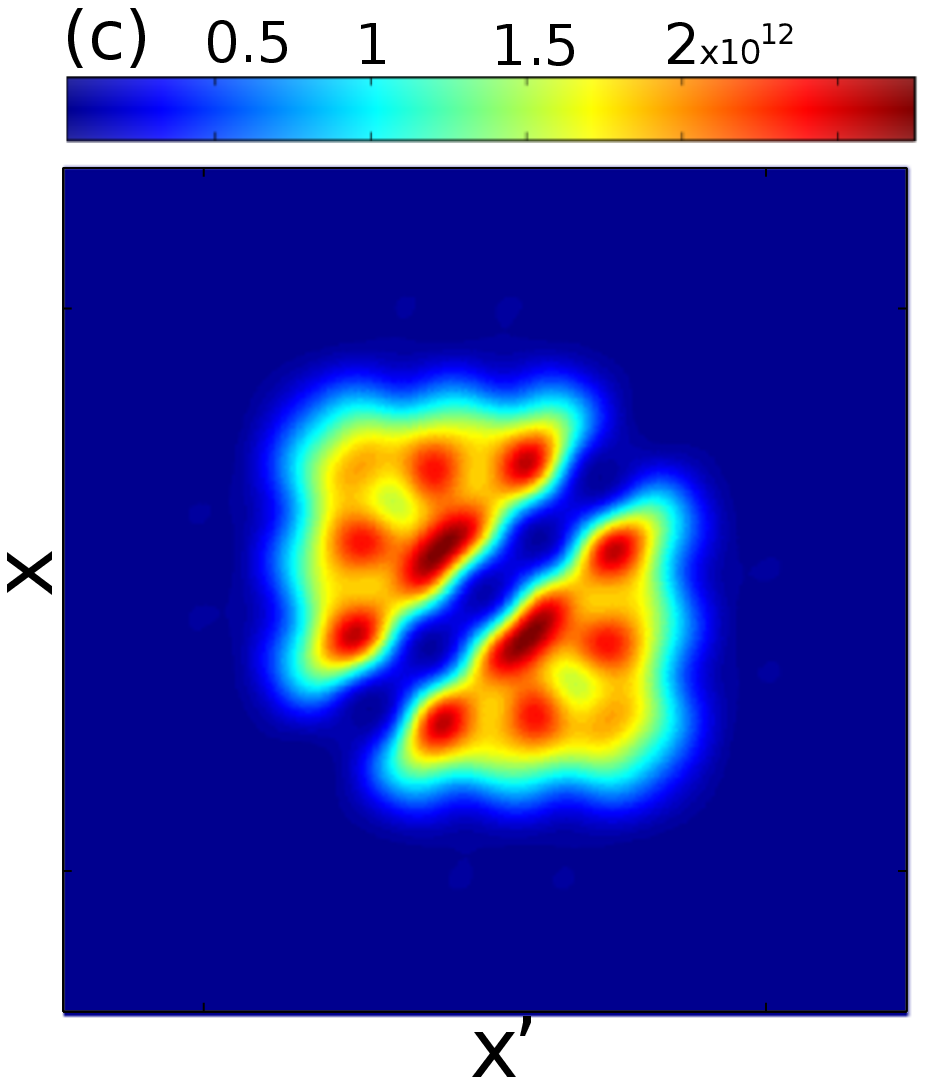} & \includegraphics[scale=0.35]{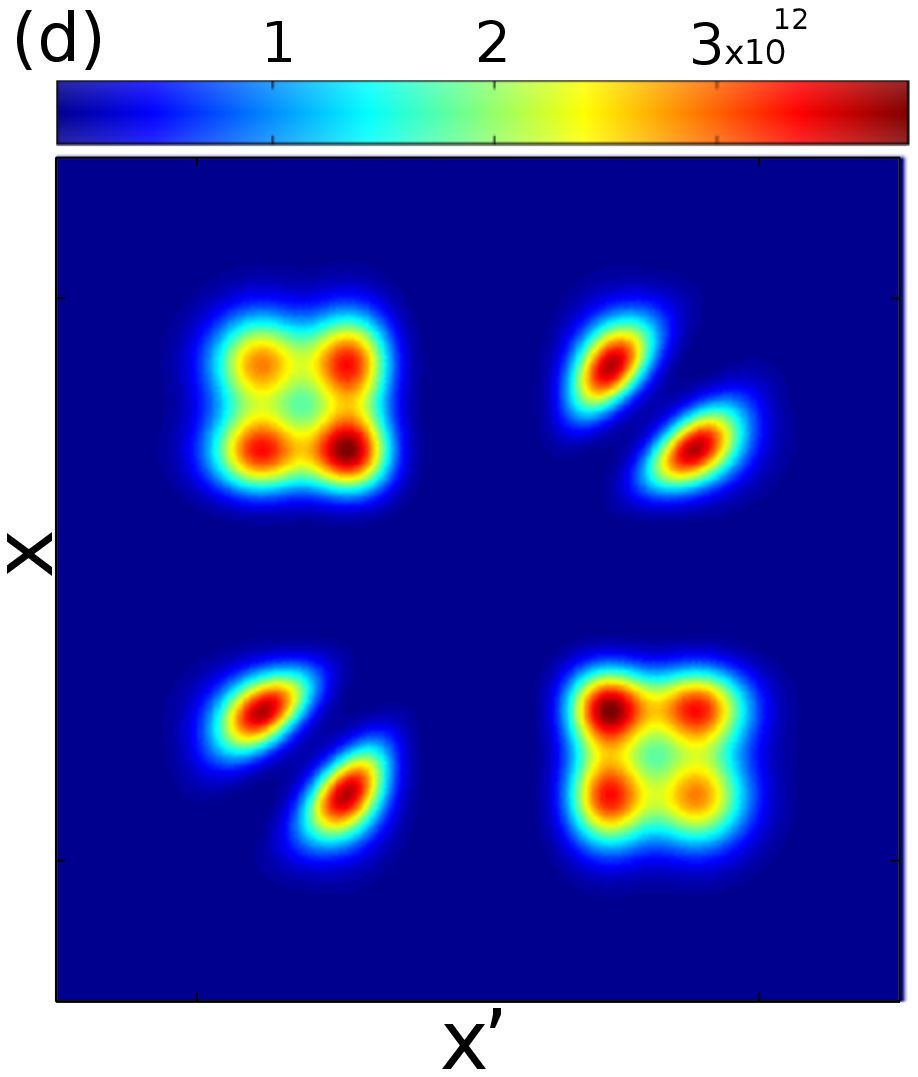}
\end{tabular}
\vspace{-0.4cm}
\caption{\label{fig:Fig3} Pair correlation functions $g^{(2)}(x,x')$
  spanning an area of $6\mu m\times 6\mu m $. The top row shows the
  BEC-BEC cases using the same parameters as in Fig.~\ref{fig:Fig1}
  for the situation (a) without and (b) with coupling between
  species. The lower row shows the BEC-GP regime analogous to the
  situation in Fig. \ref{fig:Fig2} for the situation (c) without and
  (d) with coupling between the species.}\vspace{-0.5cm}
\end{figure}

Let us finally discuss the pair correlation functions
$g^{(2)}(x,x')=\langle \hat\Psi^\dagger(x) \hat\Psi^\dagger(x')
\hat\Psi(x')\hat\Psi(x)\rangle$, with $\hat{\Psi}(x)$ being the field
operator, using again the microscopic model. These correlations give
the probability of finding an atom at position $x$ once another atom
has been measured at $x'$.  The BEC-BEC case with no interaction
between species (corresponding to the upper row of
Fig.~\ref{fig:Fig1}) is shown in Fig.~\ref{fig:Fig3}(a) and the
expected Gaussian profile is obtained. In the phase separation regime
(see Fig.~\ref{fig:Fig3}(b)) we find that the two approximately
Gaussian parts of the density profile of the system (shown in the
lower row in Fig.~\ref{fig:Fig1}) are highly correlated .  For the
BEC-TG case with no inter-species interaction (corresponding to the
upper row in Fig.~\ref{fig:Fig2}), as shown in Fig.~\ref{fig:Fig3}(c),
we recover the well-known hard-sphere behaviour that no two atoms can
be found at the same point in space, which persists in the phase
separated limit for the individual peaks (see
Fig.~\ref{fig:Fig3}(d)). The correlations between the two peaks are
also still visible, however they show a more complicated structure due
to the inter-species interaction.

In conclusion, we have presented a microscopic and a semiclassical
model to describe a two-component Bose gas at ultralow temperatures in
the one-dimensional limit. Since we allow for one component to be in
any correlation regime, our work generalises the known models for
interpenetrating Bose-Einstein condensates to include a significantly
larger group of systems, which are currently about the become
experimentally available.

From our microscopic model we have, as a first example, derived a
criterion for phase separation, which extends the well-known mean
field result to the mesoscopic limit. We have also presented a
semiclassical description of a mixture of ultracold bosons when one of
them is highly interacting, and found that the nonlinear coupling term
between the species depends on the interaction strength. Finally, we
have shown that even when the atoms are phase separated, they are
still strongly correlated among themselves. Interesting questions that
can be approached with this model include the study of quantum
correlations between a system and an environment in fundamental
settings as well as using one component (one matter wave) to engineer
the state of the second matter wave.

\begin{acknowledgments}
  This project was support by Science Foundation Ireland under Project
  No.~10/IN.1/I2979. 
\end{acknowledgments}

\end{document}